\documentclass[twocolumn,showpacs,preprintnumbers,amsmath,amssymb]{revtex4}

\usepackage{graphicx}
\usepackage{dcolumn}
\usepackage{bm}

\def\lapproxeq{\lower .7ex\hbox{$\;\stackrel{\textstyle
<}{\sim}\;$}}
\def\gapproxeq{\lower .7ex\hbox{$\;\stackrel{\textstyle
>}{\sim}\;$}}

\begin{document}

\hyphenation{Es-ta-du-al}

\preprint{BA-03-23}

\title{Influence of shower fluctuations and primary composition\\ 
on studies of the shower longitudinal development}

\author{Jaime Alvarez-Mu\~niz}
\email{jaime@fpaxp1.usc.es}
\altaffiliation[Now at ]{Dept. de F\'\i sica de Part\'\i culas, 
Facultade de F\'\i sica, Universidade de Santiago de Compostela,
15706 Santiago de Compostela, A Coru\~na, Spain}
\author{Ralph Engel}
\altaffiliation[Now at ]{Forschungszentrum Karlsruhe, Institut f\"ur
Kernphysik, Postfach 3640, 76021 Karlsruhe, Germany}
\author{T.K. Gaisser}
\author{Jeferson A. Ortiz}
\altaffiliation[Now at ]{
Instituto de Astronomia,
Geof\'{\i}sica e Ci\^encias Atmosf\'ericas, Universidade de S\~ao
Paulo, Caixa Postal 3386, S\~ao Paulo, SP 01060-970, Brasil
}
\author{Todor Stanev}
\affiliation{Bartol Research Institute,\\
University of Delaware, Newark, Delaware 19716, U.S.A.\\
}


\begin{abstract}
We study the influence of shower fluctuations, and the possible presence of 
different nuclear species in the primary cosmic ray spectrum, on the 
experimental determination of both shower energy and the proton air 
inelastic cross section from studies of the longitudinal development 
of atmospheric showers in fluorescence experiments. We investigate 
the potential of track length integral and shower size at maximum as
estimators of shower energy.  We find that at very high energy
 ($\sim 10^{19}-10^{20}$ eV) the error of the total energy assignment
 is dominated by the dependence on the hadronic interaction model,
 and is of the order of 5\%. At lower energy  ($\sim 10^{17}-10^{18}$ eV),
 the uncertainty of the energy determination due to the limited knowledge of
 the primary cosmic ray composition is more important.
 The distribution of shower maximum, $X_{\rm max}$, is discussed
 as a measure of the proton-air cross section. Uncertainties
 in a possible experimental measurement of this cross section
 introduced by intrinsic shower fluctuations, the model of
 hadronic interactions, and the unknown mixture of primary nuclei
 in the cosmic radiation are numerically evaluated.
\end{abstract}



\maketitle


\section{\label{introduction}Introduction}

The fluorescence technique of ultra high energy cosmic ray (UHECR) 
detection was first explored in the pioneering  
Fly's Eye detector \cite{Baltrusaitis88}, and is currently 
being used in its 
successor, the high resolution Fly's Eye (HiRes) \cite{Abu-Zayyad00}, 
as well as in
the Pierre Auger Observatory \cite{Cronin95} that is currently under
construction. 
The underlying idea is the detection of atmospheric nitrogen fluorescence 
light induced by the passage of charged particles through 
the atmosphere. The number of charged particles at depth $X$ in the 
atmosphere, $N(X)$, i.e. the longitudinal shower profile, 
can be extracted from data because $N(X)$ is to a good approximation
proportional to the amount of emitted fluorescence light. 
In this approximation, the total energy that
goes into electrons and positrons (the electromagnetic energy
$E_{\rm em}$ from now on) is obtained by integration of the
shower longitudinal profile \cite{Song00}
\begin{equation}
E_{\rm em}=\alpha_{\rm eff} \int_0^{\infty} N(X)dX
\label{eq:Eem}
\end{equation}  
where $\alpha_{\rm eff}$ is the  
average (effective) ionization loss rate which is usually taken as a constant
over the entire shower and is given by $\sim 2.19~{\rm MeV/g~cm^{-2}}$ 
\cite{Baltrusaitis85,Song00}. 

The integral on the right hand side of Eq.~(\ref{eq:Eem}) represents the 
total track length of all charged particles in the shower projected onto 
the shower axis.
Electrons and positrons constitute the 
bulk of the charged particles in a shower and contribute
most to the production of fluorescence light.  In the following 
we neglect the contribution of muons
and other charged particles to the production of fluorescence light,
which is of the order of 2\% (see discussion in \cite{Risse03b}).


It is generally
assumed that the fluorescence rate is proportional to the ionization
energy loss rate $dE/dX$, although this has been experimentally proved only 
to some extent \cite{Kakimoto96,Nagano:2003zn}. Consequently in order
to estimate shower energy, there is in principle no need
to convert the measured fluorescence intensity  
first to a particle number, and then relate
the total track length to the energy of the shower through
Eq.~(\ref{eq:Eem}). 
 The total ionization energy deposit can instead be obtained from the 
fluorescence intensity and can be used directly as an energy 
estimate \cite{Dawson02}.
However, as long as the lateral spread of shower particles is correctly
accounted for \cite{Alvarez-Muniz03}, the conversion of fluorescence light 
intensity to number of particles and then to energy through
Eq.~(\ref{eq:Eem})
does not lead in principle to observable errors mainly for two reasons: 
Firstly the ionization energy deposit depends only weakly on the particle 
energy, and secondly the shape of the energy spectrum of particles in an 
air shower changes only slowly with the traversed depth. Only in the 
very early evolution stage of
a shower is the particle energy spectrum significantly harder than that at the
shower maximum. The corresponding energy deposit is higher by up to a
factor of 1.5, but due to the small number of particles, the
resulting error in the energy estimation is negligible \cite{Risse02a}.  

There are several additional factors, such as air pressure, density and
humidity, that influence the relation of the
fluorescence intensity to energy deposit and particle numbers. The
discussion of these aspects, including the conversion of the observed
light curve to a longitudinal shower profile are beyond the scope of this
work.

In this article we investigate the longitudinal shower profile as
an experiment-independent quantity and study its relation to the energy
and mass of the primary particle. In Sec.~\ref{sec:energy} the 
track length integral and the particle number at shower maximum are compared as
energy estimators under the assumption of an unknown cosmic ray
composition above $10^{17}$ eV. The model and mass dependence 
of the invisible energy
carried by neutrinos and energetic muons is calculated for the QGSjet
\cite{qgsjet} and
SIBYLL \cite{Engel99,Fletcher94} 
models of hadronic interactions. The mean position of the shower
maximum, $X_{\rm max}$, and its distribution is discussed as a measure of
the primary cosmic ray composition and proton-air cross section in
Sec.~\ref{sec:xmax}. A summary and conclusions are given in
Sec.~\ref{sec:dis}.


\section{Shower track length and energy reconstruction\label{sec:energy}}

Using the particle number as primary observable, an 
accurate Monte Carlo calculation of the electromagnetic energy
in a shower requires that all the contributions to the track length
due to electrons and positrons are properly accounted for. This
implies computing the track length of electrons of a very wide energy
range, including very small energies,
since electrons with kinetic energy below 0.1 MeV 
typically account for $\sim10\%$ of the electromagnetic energy \cite{Song00}. 
A part of the energy of the primary particle is not converted to
electromagnetic particles and hence not deposited as ionization energy
in the atmosphere. High-energy muons deposit only a small
fraction of their energy in the atmosphere and
neutrinos escape detection completely. Therefore experiments have to
correct for the ``unseen'' energy to estimate the primary particle energy.
In a Monte Carlo simulation, once 
$E_{\rm em}$ is determined, the ``unseen'' energy can be calculated on a
shower-by-shower basis as the difference to the primary particle energy.

In this section we calculate the fraction of shower
energy that is not detected by a fluorescence experiment 
and we study its dependence on shower energy, on the 
primary nucleus that initiates the shower, and on the hadronic
interaction model needed to extrapolate to unmeasured regions
of the phase space of the primary nucleus-air collision. We compare 
the resolution achieved with two different estimators of shower energy, 
namely the track length integral and the number of particles at shower
maximum. For this purpose we simulate large samples of showers and 
extract their longitudinal profile. 

We use a fast hybrid simulation program~\cite{Alvarez02} 
that allows the simulation of 
longitudinal shower profiles of electrons and muons. 
The hybrid method consists of calculating shower
observables by a direct simulation of the initial part of
the shower, tracking all particles of energy above $E_{\rm thr}=0.01~E$.
Parameterizations of presimulated showers for all subthreshold particles are
then superimposed after their first interaction point is
sampled. The sub-showers are described with parameterizations
that give the correct average behavior, and at the same time
describe the fluctuations in shower development of both electrons
and muons. 

Electromagnetic cascades are simulated with a full-screening
electromagnetic Monte Carlo at high energy and at low energy
the longitudinal development of 
electromagnetic sub-showers is calculated using Greisen's
parameterization \cite{Rossi41a}. The numerical approximation as 
given in \cite{Greisen56} is applied with electron-induced showers being
shifted by 0.8 radiation lengths \cite{Gaisser97}.
Greisen's formula is a good approximation to the numerical solution of
the cascade equations with vanishing low-energy cutoff.

In order to apply Eq.~(\ref{eq:Eem}) to estimate shower energy from 
the number of particles given by our hybrid approach, 
the factor $\alpha_{\rm eff}$ in Eq.~(\ref{eq:Eem}) must 
be determined for our approximation of electromagnetic showers. 
We normalize the track length predicted by 
our hybrid method to the electromagnetic energy in photon 
initiated showers of energy $E_{\rm em}$, 
turning photoproduction interactions artificially off to avoid that
a fraction of $E_{\rm em}$ goes into a muonic and neutrino component,
\begin{equation}
\alpha_{\rm eff}=\frac{E_{\rm em}}{\int N(X) dX} \ .
\end{equation}
We obtain the numerical value $\alpha_{\rm eff}=2.32~{\rm MeV/g~cm^{-2}}$,
which we will use throughout this paper. It is important to realize that
$\alpha_{\rm eff}$ depends on the treatment of low energy 
particles due to the different kinetic energy thresholds of the 
Monte Carlo simulations, and therefore our result cannot be 
directly compared to 
the numerical value obtained by Song et al. 
($\alpha_{\rm eff}=2.19~{\rm MeV/g~cm^{-2}}$)~\cite{Song00} 
who performed a CORSIKA \cite{Heck98a} simulation with the threshold of
100 keV. Song et al. estimated that about 10\% of the electromagnetic energy is
carried by particles of kinetic energy less than 100 keV, which are
neither included in the track length simulation nor in the calculation of
$\alpha_{\rm eff}$. The value 
of $2.42~{\rm MeV/g~cm^{-2}}$
found by Risse and Heck (Fig.~7 in \cite{Risse02a}) is not in
contradiction with our result as it refers to a simulation threshold 
of 250 keV. In contrast to \cite{Song00} the analysis in \cite{Risse02a}
defines $\alpha_{\rm eff}$ as the proportionality constant between the
projected track length of all particles above simulation threshold and
the total calorimetric energy, including the expected energy deposit of
the particles falling below the simulation threshold.


\subsection{Unseen energy in hadron-induced atmospheric showers}

We turn now to the study of the unseen energy in nucleus-induced 
showers. 
Using the hybrid method we have simulated showers initiated by 
protons, helium, carbon and iron at zenith angle $\theta=45^\circ$, 
down to the approximate observation 
level of the HiRes and Auger experiments corresponding to a 
vertical depth of $X_{\rm v}=870~{\rm g/cm^2}$. We have performed
the simulations using 
two hadronic interaction models  SIBYLL 2.1 \cite{Engel01} 
and QGSjet01 \cite{qgsjet}, with the aim to   
study the influence of the model predictions on the 
amount of unseen energy in the shower.  

We calculate the total track length by performing the integral in depth of 
the longitudinal profile generated by the hybrid method. Similar to the
method applied by air-fluorescence experiments the simulated curve is
fit by the Gaisser-Hillas function~\cite{Gaisserbook}
\vspace{-0.1cm}
\begin{eqnarray}
N_{\rm GH}(X)=N_{\rm max}~
\left( {X-X_0 \over X_{\rm max}-X_0} \right)^{
(X_{\rm max}-X_0)/\lambda}\nonumber\\
\times~{\rm exp}
\left[-{(X-X_{\rm max})\over\lambda}\right],
\label{eq:GH}
\end{eqnarray}
where $X_0$ is the depth of
the first interaction, $X_{\rm max}$ is the depth at
which the number of electrons in the shower is maximal,
and $N_{\rm max}$ is the number of electrons at maximum.
First the position of the shower maximum and the
particle number at maximum are found by a polynomial fit to the shower
profile near its maximum.
In a subsequent fit the parameters $\lambda$ and $X_0$ are determined. 
The unseen energy $E_{\rm u}$ in a shower of energy $E$ then follows
from
\begin{equation}
E_{\rm u}=E-E_{\rm em}=E-\alpha_{\rm eff}\int_0^{\infty}N_{\rm GH}(X)dX
\ .
\end{equation}

\begin{figure}
\centerline{\includegraphics[width=8.5cm]{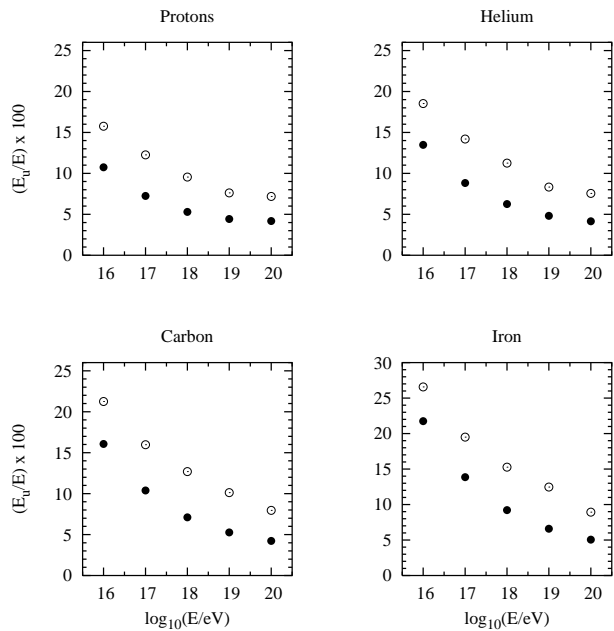}}
\caption{Fraction of the total shower energy that is not seen 
by a fluorescence detector in showers initiated by proton, helium,
carbon and
iron as a function of shower energy $E$. 
5,000 showers at zenith angle
$\theta=45^\circ$ were simulated for each hadronic
interaction model: SIBYLL 2.1 (bold circles) and QGSjet01 (open circles).
}
\label{fig:Einv}
\end{figure}

In figure~\ref{fig:Einv} we plot the mean unseen energy as obtained 
in showers initiated by different nuclei and compare the 
predictions of the interaction models SIBYLL 2.1 and QGSjet01.
The energy that is transferred
to muons and neutrinos decreases with shower energy for all nuclei in 
both models. The main reason for this behavior is that as shower 
energy increases the average energy of charged 
pions increases as well, and in turn their probability of decaying 
into muons and neutrinos diminishes. At fixed energy the unseen 
energy is larger in showers initiated by heavy nuclei than in those 
induced by light nuclei. This can be understood on the basis of the
superposition model in which a shower induced by a nucleus of $A$
nucleons and energy $E$ is considered as $A$ independent proton showers of 
energy $E/A$, in each of which the fraction of unseen energy is larger 
than in a proton shower of energy $E$ (Fig.~\ref{fig:Einv}). 
In fact, the muon number in a proton shower scales
as $N_\mu \sim E^\alpha$ with $\alpha = 0.86 \dots 0.92$ \cite{Alvarez02}, 
and applying the superposition model it can be easily seen that 
an iron shower has about 1.5 times more muons than a proton shower of 
the same energy.

The unseen energy reaches
almost a constant value at shower energy above $10^{19}-10^{20}$ eV, and
is fairly insensitive to composition in this 
energy region, the relative difference between the unseen
energy fractions among the different nuclei being $\sim 10\%$. 
Interestingly the difference between the two model predictions
is much bigger and essentially stays the same.
At ultra-high energy the
uncertainty in the missing energy assignment is dominated by the
model dependence and not by an unknown primary composition.

The unseen energy predicted
by QGSjet01 is consistently larger than the corresponding energy 
predicted by SIBYLL 2.1, the reason 
being that the muon multiplicity ($E_\mu > 0.3~{\rm GeV}$) 
in QGSjet01 is larger than in SIBYLL 2.1. For example, QGSjet predicts 
about 35\% more muons than SIBYLL 2.1 
in proton-induced showers at $10^{20}$ eV \cite{Alvarez02}. This
difference arises mainly due to the different multiplicities of 
secondaries predicted by the two models. 
The relative difference between the saturation values of the unseen energy 
predicted by SIBYLL and QGSjet is $\sim 50\%$. This translates to a
$5\%$ uncertainty for the total energy estimate.

Our results for the unseen energy fraction are similar to those obtained in
\cite{Song00} and recently in \cite{Barbosa03}. For example, Song et al. 
find a fraction of unseen energy of about 7\% (10\%) at $E=10^{20}$ eV 
for proton (iron) showers and QGSjet98 \cite{Song00}. 
Barbosa et al. have also used
SIBYLL 2.1 in their simulations. They obtain an unseen energy fraction
of about 5\% and 8\% for proton and iron showers at $10^{20}$ eV
simulated with the SIBYLL model~\cite{Barbosa03}. There are differences
of the order of 1-3\%
between these calculations and our results which we
attribute to the different low-energy interaction models used for the
simulations, the approximative character of Greisen's parametrization
for low-energy electromagnetic sub-showers, and different methods
 of calculating the
track length integral. The latter involves extrapolating the electromagnetic
shower component to larger atmospheric depths.


\subsection{Shower energy reconstruction from the longitudinal shower profile}

There are several methods
to reconstruct the primary energy 
of an observed air shower profile experimentally. 
Not only the total shower track length in the atmosphere in 
Eq.~(\ref{eq:Eem}) but also the number of particles at shower 
maximum can serve as an estimator of the shower energy.  
The latter one is of particular interest for nearly vertical showers
where only the first part of the shower can be observed and the
uncertainty in the calculation of the total track length could be 
dominated by the extrapolation
of the shower profile into the unobserved region.
In this section we study the energy resolution that is achieved 
with these two methods.

As previously discussed,
shower energy can be calculated from the measured track length.
The procedure is to fit a Gaisser-Hillas or other function to 
the observed longitudinal profile \cite{Abu-Zayyad00b}, integrate it 
to obtain the total track length, extract the energy
that goes into the electromagnetic component and correct it 
for the unseen energy that is estimated by Monte Carlo simulations. 

The procedure above is subject to uncertainties because, usually, 
a previously calculated mean value of the missing energy, averaged 
over many simulated showers, is used
to determine the energy of each individual event. Moreover, due to fluctuations
in the shower longitudinal profile the experiments are unable to 
determine the type of primary nucleus that initiated the shower on
an event-by-event basis. In consequence a correction 
for missing energy averaged over different primaries must be used. 
In addition the correction for unseen energy depends on the hadronic 
interaction model used to perform the simulations. 

We estimate the energy of the shower ($E_t$) from the track length obtained 
as indicated above in the following way, 
\begin{equation}
E_t=\langle f_u \rangle E_{\rm em}=
\langle f_u \rangle \alpha_{\rm eff} \int_0^{\infty} N_{\rm GH}(X)dX 
\label{eq:E-track}
\end{equation} 
where $\langle f_u \rangle$ is the average value of the correction for 
the energy not seen. The value of $\langle f_u \rangle$ 
depends on $E_{\rm em}$ and 
is for simplicity taken as the arithmetic mean over a uniform four-component
mass composition, i.e. $\langle f_u \rangle=(\sum_i f_u^i)/4$
where $f_u^i=E_{\rm em}/E$ are the corrections for unseen energy in 
showers initiated by different primary nuclei. The index
$i$ corresponds to proton, helium, carbon and iron induced showers. 
For fixed primary particle type, 
$f_u^i$ was obtained as the average value of $E_{\rm em}/E$ in 5,000 
simulated showers. 

Alternatively, shower energy can also be estimated from the size of the 
electron distribution at shower maximum $N_{\rm max}$. 
The relation between $N_{\rm max}$ and energy can be expressed as 
$E=g N_{\rm max}$ where $g$ must be obtained by Monte Carlo 
simulations. As before, the determination of shower energy 
is subject to uncertainties because $g$ has to 
be averaged over many showers and different primaries, and is also model
dependent.  Then the estimated energy of a shower ($E_N$) follows from 
$N_{\rm max}$ through the equation,
\begin{equation}
E_N=\langle g \rangle N_{\rm max}\ ,
\label{eq:E-Nmax}
\end{equation}
where $\langle g \rangle=(\sum_i g_i)/4$ and the index $i$ corresponds
again to protons, helium, carbon and iron. The $g_i$ values
were obtained as the average of $E/N_{\rm max}$ in 5,000
simulated showers fixing the type of primary particle.

\begin{figure}
\centerline{\includegraphics[width=7cm]{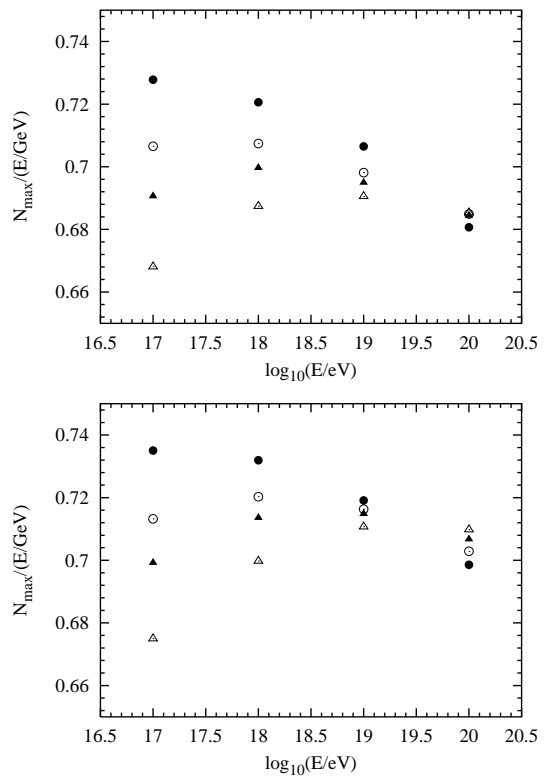}}
\caption{Average shower size at maximum normalized to shower energy
in GeV in showers initiated by 
proton (bold circles), helium (open circles), carbon (bold triangles) and
iron (open triangles) as a function of shower energy $E$.
5,000 showers at zenith angle
$\theta=45^\circ$ were simulated for each hadronic
interaction model: SIBYLL 2.1 (bottom panel) and QGSjet01 (top panel).
}
\label{fig:smax}
\end{figure}

Figure~\ref{fig:smax} shows the average shower size at maximum as a function
of energy in showers initiated by different nuclei. Each point represents
an average over 5,000 showers. The results of SIBYLL 2.1 and QGSjet01
are presented. It is interesting to see how the dependence of 
$N_{\rm max}$ on composition changes with shower energy. 
In the energy region of $10^{19}$ to $10^{20}$ eV the value of $N_{\rm
max}$ is rather composition-independent, making it a good energy
estimator. However, the mass dependence is of the order of 10\% in
the energy range between $10^{17}$ and $10^{18}$ eV. The hadronic
interaction model dependence for fixed energy and type of primary
is remarkably small $\sim 1-2\%$. 

\begin{figure}
\centerline{\includegraphics[width=8.5cm]{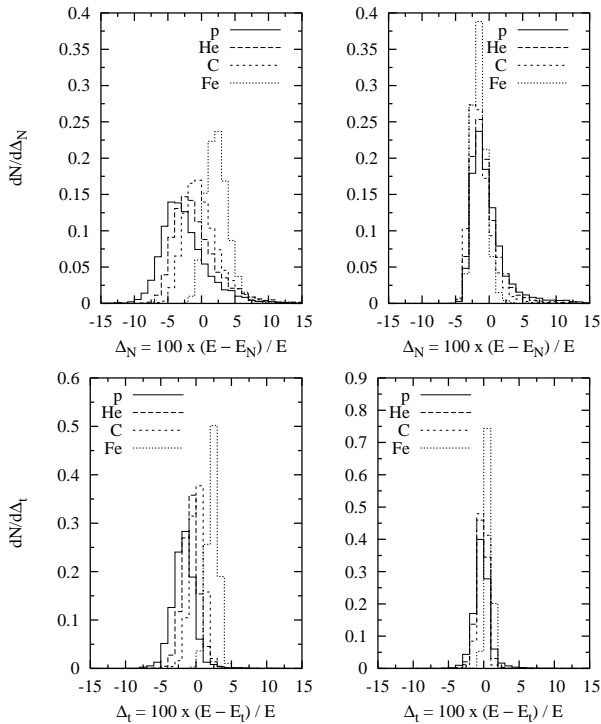}}
\caption{Distributions of the error in shower energy determination for
two energy estimators. The results are shown
for different nuclei and for SIBYLL 2.1. 
Left panels $E=10^{18}$ eV, right panels $E=10^{20}$ eV. Top
panels: $N_{\rm max}$ is used as energy estimator. 
Bottom panels: the energy estimator is the track length integral.}
\label{fig:Eres_SI21}
\end{figure}

\begin{figure}
\centerline{\includegraphics[width=8.5cm]{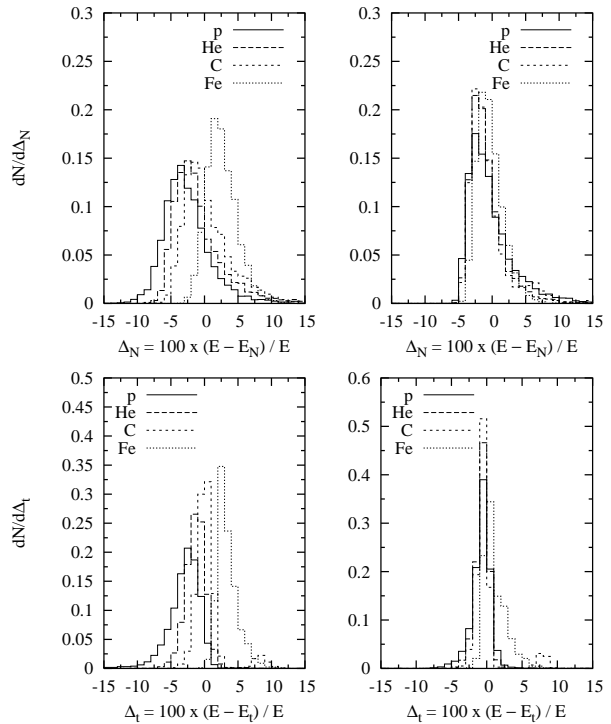}}
\caption{Distributions of the error in shower energy determination for
two energy estimators. The results are shown
for different nuclei and for QGSjet01.
Left panels $E=10^{18}$ eV, right panels $E=10^{20}$ eV. Top
panels: $N_{\rm max}$ is used as energy estimator in the top panels.
Bottom panels: the energy estimator is the track length integral.}
\label{fig:Eres_QG01}
\end{figure}

For an experiment measuring showers with a steep energy spectrum the
knowledge of the event-by-event correlation, i.e. the energy resolution, 
is of great importance. In Fig.~\ref{fig:Eres_SI21} we show 
the resolution in energy achieved with $E_t$ in Eq.~(\ref{eq:E-track}),
and $E_N$ in Eq.~(\ref{eq:E-Nmax}), for SIBYLL 2.1 at two representative 
shower energies. 
We plot the relative difference in percentage between the estimated 
energy ($E_t$ or $E_N$) and the actual shower energy.
Fig.~\ref{fig:Eres_QG01} shows the corresponding energy resolution 
for QGSjet01.    
The mean values and standard deviations of the proton and iron 
distributions shown in Figs.~\ref{fig:Eres_SI21} and \ref{fig:Eres_QG01} 
are given in tables~\ref{tab:Eres_p} and \ref{tab:Eres_Fe}. 

At energies above $10^{19}$ eV, the energy resolution achieved with 
$N_{\rm max}$ is similar to the one obtained when the track length 
is used as energy estimator.  
However the error in $E$ with $N_{\rm max}$ is more asymmetric
than the corresponding one for the track length, due to the intrinsic
asymmetry of the distribution of $N_{\rm max}$ at $E=10^{19}-10^{20}$ eV 
(see for instance Fig. 11 in reference~\cite{Alvarez02}). The asymmetry
is less pronounced in the case of heavy nuclei.  

Both energy estimators produce an energy resolution which is not 
strongly dependent on composition above $E=10^{19}$ eV.
This can be easily understood from 
Figs.~\ref{fig:Einv} and \ref{fig:smax} in which 
it can be seen that at $E>10^{19}$ eV both the fraction of unseen energy in the
shower and $N_{\rm max}$ are weakly dependent on the type of primary nucleus.
The dependence on composition is more pronounced in the energy 
range $10^{17}-10^{18}$ eV, on average a $\sim 5-7\%$ uncertainty in 
the energy determination with both methods is introduced if the nature 
of the primary is unknown. It is interesting to notice that both methods
tend to underestimate the energy for proton primaries and overestimate it
for iron primaries due to the assumption of a uniform four-component
composition in all the energy range above $10^{17}$ eV.
The systematic uncertainty introduced by 
the hadronic interaction model when using $N_{\rm max}$ as estimator 
is about $1\%$, comparable to the one in the case of the 
track length integral method.

In summary, $N_{\rm max}$ is a remarkably good energy estimator 
that may replace the total track length at energies above $10^{19}$ eV 
if care is taken for the asymmetric errors in the distributions. 
An $N_{\rm max}$-based energy determination may be superior to the 
track length method when only a small portion
of the shower around the maximum is detected, for instance 
due to the limited sensitivity or acceptance of the fluorescence detector.


\begin{table*}
\caption{Mean value and standard deviation of the distributions of
the error corresponding to the histograms for protons in
Figs.~\ref{fig:Eres_SI21}
and \ref{fig:Eres_QG01}.
\label{tab:Eres_p}}
\renewcommand{\arraystretch}{1.5}
\begin{tabular}{ c c c c c c c c c } \hline\hline
Model & \multicolumn{4}{c }{SIBYLL 2.1} & \multicolumn{4}{c }{QGSjet01} \\


~$\log_{10}$(E/eV)~&~~$\langle \Delta_t \rangle$~~&~$\sigma({\Delta_t})$~~&~~$\langle \Delta_N \rangle$~~&~~$\sigma({\Delta_N})$~~&~~$\langle \Delta_t \rangle$~~&~~$\sigma({\Delta_t})$~~&~~$\langle \Delta_N \rangle$~~&~~$\sigma({\Delta_N})$~~\\ \hline

~$17.0$~&~~$-3.2$~~&~~$6.1$~~&~~$-4.4$~~&~~$5.4$~~&~~$-3.8$~~&~~$6.8$~~&~~$-4.6$~~&~~$5.1$~~\\

~$18.0$~&~~$-1.8$~~&~~$2.6$~~&~~$-2.4$~~&~~$3.8$~~&~~$-3.0$~~&~~$6.7$~~&~~$-2.5$~~&~~$3.8$~~\\

~$19.0$~&~~$-0.9$~~&~~$1.7$~~&~~$-0.3$~~&~~$3.0$~~&~~$-2.3$~~&~~$2.0$~~&~~$-1.6$~~&~~$3.1$~~\\

~$20.0$~&~~$-0.2$~~&~~$2.0$~~&~~$-0.1$~~&~~$3.2$~~&~~$-0.8$~~&~~$2.2$~~&~~$-0.1$~~&~~$3.8$~~\\ \hline\hline

\end{tabular}
\end{table*}


\begin{table*}
\caption{Mean value and standard deviation of the distributions of
the error corresponding to the histograms for iron in Figs.~\ref{fig:Eres_SI21}
and \ref{fig:Eres_QG01}.
\label{tab:Eres_Fe}}
\renewcommand{\arraystretch}{1.5}
\begin{tabular}{ c c c c c c c c c } \hline\hline
Model & \multicolumn{4}{c }{SIBYLL 2.1} & \multicolumn{4}{c }{QGSjet01} \\


~~$\log_{10}$(E/eV)~~&~~$\langle \Delta_t \rangle$~~&~~$\sigma({\Delta_t})$~~&~~$\langle \Delta_N \rangle$~&~~$\sigma({\Delta_N})$~~&~~$\langle \Delta_t \rangle$~&~~$\sigma({\Delta_t})$~~&~~$\langle \Delta_N \rangle$~~&~~$\sigma({\Delta_N})$~~\\ \hline 

~$17.0$~&~~$4.1$~~&~~$1.6$~~&~~$4.3$~~&~~$2.2$~~&~~$4.1$~~&~~$2.3$~~&~~$4.3$~~&~~$2.5$~~\\

~$18.0$~&~~$2.4$~~&~~$0.7$~~&~~$2.3$~~&~~$1.7$~~&~~$3.4$~~&~~$2.5$~~&~~$2.6$~~&~~$2.2$~~\\

~$19.0$~&~~$1.4$~~&~~$0.3$~~&~~$1.1$~~&~~$1.4$~~&~~$3.1$~~&~~$3.2$~~&~~$1.0$~~&~~$2.2$~~\\

~$20.0$~&~~$0.6$~~&~~$0.2$~~&~~$-1.4$~~&~~$1.1$~~&~~$1.1$~~&~~$2.7$~~&~~$-0.3$~~&~~$2.0$~~\\ \hline\hline

\end{tabular}
\end{table*}


\section{Shower longitudinal development and p-air cross section 
determination\label{sec:xmax}}

The most obvious, experiment-independent observable characterizing the
longitudinal shower profile is the depth of maximum, $X_{\rm max}$. The
mean depth of maximum is a good measure of the composition in units of
the mean logarithmic mass. Indeed, within the superposition model one
expects for a primary particle with mass number $A$
\begin{equation}
\langle X^{(A)}_{\rm max} \rangle = D_e \ln(E/A) = 
\langle X_{\rm max}^{(p)}\rangle - D_e \ln A,
\end{equation}
with $D_e$ being the elongation rate, a weakly energy dependent 
parameter. The elongation rate reflects 
features of high-energy hadron production, see for example
\cite{Linsley80a,Alvarez02}, making the interpretation of measurements
model-dependent.
However not only the mean depth of maximum carries important information 
about both the primary cosmic ray composition and the features of the 
hadronic interactions, but also its distribution. A
number of analyses using $X_{\rm max}$ distributions to infer the 
primary cosmic ray chemical composition are available in
literature, for example \cite{Gaisser:1993ix,Abu-Zayyad:2000ay,HiRes03a}.
In the following we will concentrate on the $X_{\rm max}$ 
distribution of showers
in respect to the possibility and limitations of determining the
features of the hadronic interactions, in particular of measuring the 
high-energy proton-air inelastic cross section. 

The correlation of the first
interaction point with the depth of shower maximum for proton showers
was first employed by the Fly's Eye Collaboration in \cite{Baltrusaitis84}. An
analysis in the context of a mixed primary composition was done in
\cite{Gaisser:1993ix}, using the Fly's Eye data and more recently
in \cite{HiRes03b} using HiRes data.

The probability of having the first interaction point of a shower, 
$X_{\rm int}$, at a depth greater than $X$ is 
\begin{equation}
P(X_{\rm int} > X) \propto \exp(-X/\lambda_{\rm int}) ,
\label{eq:lambda_int}
\end{equation}
with the interaction length $\lambda_{\rm int} = \langle m \rangle / \sigma$. 
Here $\langle m \rangle$ is
the mean mass of the air nuclei and $\sigma$ denotes the inelastic,
particle production cross section. 
In case of a perfect correlation between
$X_{\rm max}$ and $X_{\rm int}$, i.e. in case fluctuations in 
shower development were non-existent, one could use directly the exponential
distribution of showers with large $X_{\rm max}$ to calculate $X_{\rm int}$ 
and hence the
proton-air cross section. However, intrinsic shower fluctuations
modify the relation between the depth of maximum distribution and the
interaction length. This modification is typically expressed by a factor $k$
with
\begin{equation}
P(X_{\rm max} > X) = B \exp(-X/\Lambda), \hspace*{0.5cm} \Lambda  = k
\lambda_{\rm int} \ .
\label{eq:Lambda}
\end{equation}

The factor $k$ depends mainly on the pace of energy dissipation in the early
stages of shower evolution. Concerning the models QGSjet and SIBYLL we 
have the interesting situation that, if one considers the mean depth 
of maximum, $\langle X_{\rm max} \rangle$, 
the mean inelasticity compensates to some extent
the differences in the cross sections. For example, SIBYLL predicts
the larger proton-air cross section but at the same time a smaller
inelasticity than QGSjet. In contrast, the ratio $\Lambda/\lambda_{\rm int}$, 
i.e. the $k$ factor, is sensitive to the inelasticity fluctuations. 
In general, a model with small fluctuations
in secondary particle multiplicity and inelasticity is characterized by
a smaller $k$ factor than a model with large fluctuations. Under the
assumption of similar fluctuations in multiplicity and inelasticity,
a model predicting a large average number of
secondary particles leads to smaller overall fluctuations of the
cumulative shower profile of the secondary particles and hence to a
smaller $k$ factor. Therefore
the parameter $\Lambda$ is very sensitive to the hadronic interaction
model and its measurement would allow one to draw conclusions on general
features of high-energy hadron production \cite{Block:1999ub}. 
Conversely, knowing the $k$
factor for a given model one can measure the proton-air cross section. 

\begin{figure}[!htb]
\centerline{
\includegraphics[width=7cm]{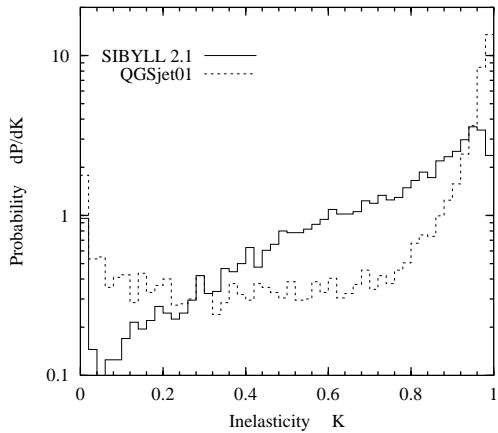}
}
\caption{
Inelasticity distribution in p-air collisions at \protect$E=10^{19}$eV.
\label{fig:inelasticity}
}
\end{figure}
In Fig.~\ref{fig:inelasticity} we show the inelasticity distribution in
proton-air interactions at $10^{19}$eV simulated with SIBYLL 2.1 
and QGSjet01.  Here
inelasticity $K$ is defined as $K = (E-E_{\rm lead})/E$ with $E_{\rm
lead}$ being the energy of the fastest baryon ($p$, $n$, or $\Lambda$)
and $E$ the projectile energy.
Although the mean inelasticities predicted by QGSjet and SIBYLL
are similar (0.77 and 0.72 respectively),
the pronounced peaks at small and large inelasticities in QGSjet events
induce somewhat larger shower-to-shower fluctuations (see below). 
The large difference in the predicted mean charged particle
multiplicities (QGSjet: 540, SIBYLL: 315 at $E=10^{19}$ eV \cite{Alvarez02}) 
is of lesser importance since most of the secondary particles are very slow.

\begin{figure}[!htb]
\centerline{
\includegraphics[width=8.5cm]{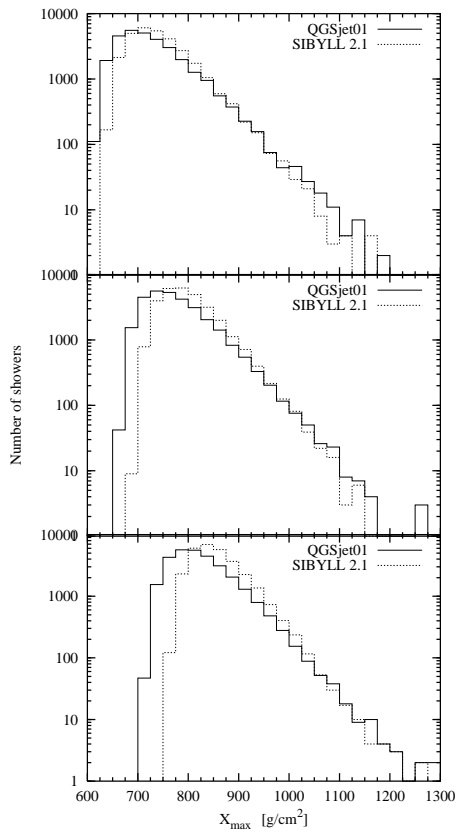}
}
\caption{
Distribution of the depth of maximum of proton-induced showers
at energies $E=10^{18},~10^{19}$ and $10^{20}$ eV (from top to 
bottom panels). 30,000 showers were simulated with QGSjet01 
(solid histogram) and SIBYLL2.1 (dotted histogram).
\label{fig:xmax-dis}
}
\end{figure}

Fig.~\ref{fig:xmax-dis} shows the $X_{\rm max}$ 
distributions for proton showers at fixed energies as predicted by
QGSjet01 and SIBYLL 2.1. The 
different slopes of the tails of the distributions 
stem from the different slopes of the exponential
first interaction probability (Eq.~(\ref{eq:lambda_int})), 
and the different intrinsic shower fluctuations predicted by the 
models. In table 
\ref{tab:slopes} we give the numerical values of the slope  
of the $X_{\rm max}$ distribution, $\Lambda$, obtained by doing an
exponential fit to the tail of the SIBYLL and QGSjet distributions
using Eq.~(\ref{eq:Lambda}). 

In the absence of internal fluctuations, all showers would  
develop through the same amount of matter, $\Delta X$, between the 
first interaction point and maximum.  As a consequence, 
a perfect correlation between $X_{\rm max}$ and 
$X_{\rm int}$ would exist, and their distributions would have exactly 
the same shape, shifted by a constant $\Delta X=X_{\rm max}-X_{\rm int}$. 
 In that case the slope of the $X_{\rm max}$ distribution $\Lambda$
 would be equal to the mean interaction length $\lambda_{\rm int}$.
 Table 
\ref{tab:slopes} compares the predictions of SIBYLL 2.1 and QGSjet01 on 
these three quantities. Also shown in the table is the standard 
deviation of the 
distribution in $\Delta X$, which gives an idea of the 
size of the fluctuations in the shower longitudinal profile. 

 The effect of fluctuations in $\Delta X$ is to broaden
 the correlation of $X_{\rm max}$ with $X_{\rm int}$ and
 to change its slope.
SIBYLL 2.1 predicts less fluctuations than QGSjet01, the difference
between the widths of the $\Delta X$ distributions 
is however fairly small $\sim~6-7~{\rm g/cm^2}$. The different 
fluctuations of the two models are also reflected 
in the larger width of the QGSjet01 $X_{\rm max}$
distribution compared to SIBYLL 2.1. 
QGSjet01 also predicts a smaller p-air cross section
(larger $\lambda_{\rm int}$)
than SIBYLL 2.1. This, added to the fact that the intrinsic fluctuations
in shower development are larger in QGSjet01,
makes the slope of the QGSjet distribution in $X_{\rm max}$ 
flatter than the SIBYLL one, as
can be seen in Fig.~\ref{fig:xmax-dis} and in table \ref{tab:slopes}.
Interestingly, the $k$ factors in SIBYLL and QGSjet01 are very similar 
within their statistical errors, somewhat larger in QGSjet01 
(table \ref{tab:slopes}). This means that the difference in the 
slopes of the $X_{\rm max}$ distributions is dominated by the different
p-air interaction lengths predicted by the models, implying that the 
larger intrinsic shower 
fluctuations of QGSjet01 play a less 
important role. This conclusion is different when the same analysis is
done for the old version
of QGSjet, namely QGSjet98. QGSjet98 predicts larger fluctuations 
than QGSjet01, mainly due to a larger diffractive cross section,
added to the fact that both versions 
have the same total and inelastic cross section and hence the same 
p-air interaction length. The larger multiplicity predicted by 
QGSjet01 further reduces the fluctuations. 
As a consequence QGSjet98 predicts  
larger $k$-factors than QGSjet01 and SIBYLL 2.1. 
Numerical values of $k$ in QGSjet98 for energies $E=10^{18}, 10^{19}$ and 
$10^{20}$ eV are $k=1.20 \pm 0.02,~1.24 \pm 0.02$ and $1.16 \pm 0.02$
respectively, the corresponding reduced $\chi^2$ are 
$\chi^2/{\rm dof}=0.56,~0.91$ and 1.18  
\footnote{The fit to the tail of the $X_{\rm max}$
distribution in order to obtain the numerical values of $k$ 
was performed using only the trailing edge of the distribution
$100~{\rm g/cm^2}$ beyond its peak.}.  

Finally, within a single model (QGSjet01 or SIBYLL 2.1), 
the $k$-factors depend very weakly on primary energy. 
This is just reflecting the weak energy dependence
of the intrinsic shower fluctuations as can be demonstrated by looking
at the values of $\sigma(\Delta X)$ in table \ref{tab:slopes}.

\begin{table*}
\caption{p-air interaction length ($\lambda_{\rm int}$), slope of the 
fitted tail of the $X_{\rm max}$ distribution ($\Lambda$), and 
standard deviation of the $X_{\rm max}$ and $\Delta X=X_{\rm max}-X_{\rm int}$ 
distributions (all in ${\rm g/cm^2}$), where $X_{\rm max}$ is the depth
of shower maximum, and $X_{\rm int}$ is the depth at which the first
p-air interaction occurs. These quantities are shown   
for different shower energies
as predicted by SIBYLL 2.1 and QGSjet01. 30,000 proton showers were simulated
to make each of the distributions. The fit to the tail of the $X_{\rm max}$ 
distribution in order to obtain the numerical values of $\Lambda$ and 
$k=\Lambda/\lambda_{\rm int}$ 
was performed using only the trailing edge of the distribution, 
$100~{\rm g/cm^2}$ beyond the peak of the distribution i.e., 
$X_{\rm max}>X_{\rm max}^{\rm peak}+100~{\rm g/cm^2}$.
Also shown is the $\chi^2/{\rm dof}$ of the fit. 
Errors, where shown, are statistical. 
\label{tab:slopes}}
\renewcommand{\arraystretch}{1.5}
\begin{tabular}{ c|c c c c c c | c c c c c c} \hline\hline
Model & \multicolumn{6}{c|}{SIBYLL 2.1} & \multicolumn{6}{c }{QGSjet01} \\

E~[eV]& $\lambda_{\rm int}$ & $\Lambda$ & $\sigma(X_{\rm max})$~&~$\sigma(\Delta X)$ & $k$ & $\chi^2/{\rm dof}$ &~ $\lambda_{\rm int}$ & $\Lambda$ & $\sigma(X_{\rm max})$~&~$\sigma(\Delta X)$ & $k$ & $\chi^2/{\rm dof}$ \\ \hline

$10^{18}$ & 43.64 & $50.63 \pm 0.70$ & 59.03 & 39.16~&~$1.15 \pm 0.03$ & 1.24 &~48.44 & $58.33 \pm 0.87$ & 66.34 & 44.94 & $1.18 \pm 0.03$ & 1.83 \\

$10^{19}$ &~39.49~&~$47.12 \pm 0.68$~&~54.74~&~37.66~&~$1.16 \pm 0.02$~&~1.08~&~44.93~&~$53.58 \pm 0.85$~&~63.32~&~44.79~&~$1.18 \pm 0.02$~&~0.90 \\

$10^{20}$ & 35.93 & $42.49 \pm 0.73$ & 51.18 & 37.56 & $1.14 \pm 0.03$ & 0.94 &~ 41.89 & $49.28 \pm 0.69$ & 59.98 & 44.51 & $1.14 \pm 0.02$ & 0.79 \\ \hline\hline

\end{tabular}
\end{table*}

\subsection{Influence of fitting range on $k$-factor determination}

\begin{figure}[!htb]
\centerline{
\includegraphics[width=10.cm]{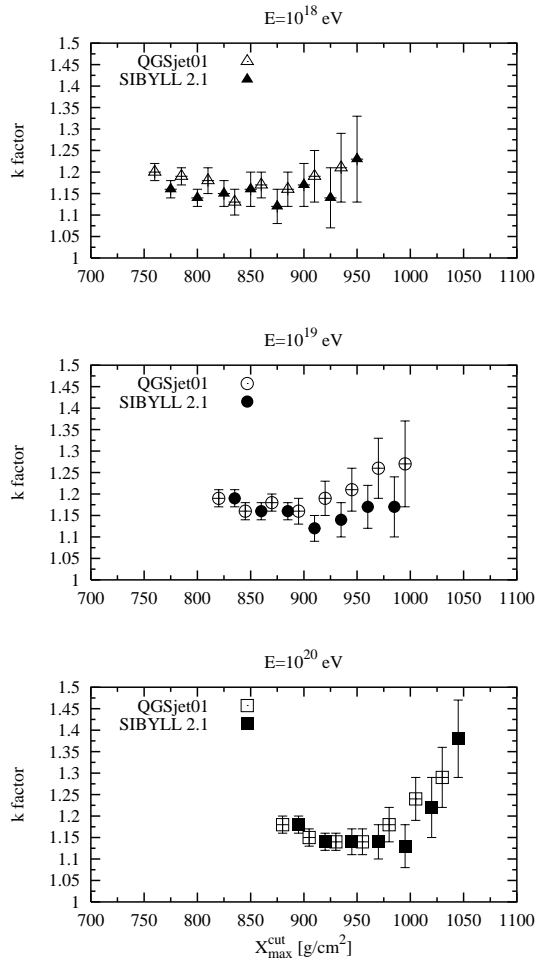}
}
\vspace*{-1cm}
\caption{
Numerical values of the $k$ factor for proton-initiated showers 
simulated with SIBYLL 2.1 (full symbols) and QGSjet01 (empty symbols) 
in dependence on the considered minimal atmospheric depth, 
$X_{\rm max}^{\rm cut}$, above which the fit to the tail of  
the distribution in $X_{\rm max}$ is performed. 30,000 showers
were simulated to make the distributions. 
\label{fig:kfactor_SI21_QG01_13}
}
\end{figure}

There are a number of complications making the measurement of the
parameter $\Lambda$ difficult. First, the $X_{\rm max}$ distribution is
not a perfect single exponential. As a consequence the $k$-factor depends on 
which part of the $X_{\rm max}$ distribution is used for fitting. 
This is illustrated in
Fig.~\ref{fig:kfactor_SI21_QG01_13} 
where a non-constant $k$ factor as a 
function of the smallest $X_{\rm max}$ considered in the fit
($X_{\rm max}^{\rm cut}$), is shown
for SIBYLL and QGSjet respectively. To test how well a single 
exponential would describe the tail of the $X_{\rm max}$ 
distributions in both models, we made fits of the $k$ factor
as function of $X_{\rm max}^{\rm cut}$, assuming $k$ being constant.
The mean values are $k=1.15 \pm 0.01,~1.16 \pm 0.01$ 
and $1.16 \pm 0.01$ at $E=10^{18},~10^{19}$ and
$10^{20}$ eV respectively for SIBYLL 2.1, the corresponding $\chi^2$ values
per degree of freedom being $\chi^2/{\rm dof}=0.25,~0.71$ and $1.6$.  
For QGSjet01 the corresponding values of $k$ and $\chi^2/{\rm dof}$ are:
$k=1.18 \pm 0.01,~1.18 \pm 0.01,~1.16 \pm 0.01$ and 
$\chi^2=0.65,~0.60,~1.48$. 

 The hypothesis of a flat behavior of $k$ versus $X_{\rm max}^{\rm cut}$ 
 is not as bad as Fig.~\ref{fig:kfactor_SI21_QG01_13} might indicate.
 The large errors of the points at large $X_{\rm max}^{\rm cut}$, 
 although they deviate most from a constant value, have the smallest
 weights in the fit and do not affect much $\chi^2/{\rm dof}$.
 The large errors stem from the lack of statistics
in the far tail of the $X_{\rm max}$ distribution  
\footnote{Despite the fact that we have
simulated 30,000 showers.}.
These errors might introduce large uncertainties in the determination
of the p-air inelastic cross section, especially if a cut at large 
$X_{\rm max}$ has to be applied in order to avoid contamination by other 
nuclear species that might be present in the primary cosmic ray spectrum. 
This last issue is the subject of subsection C below.

\subsection{Influence of resolution in $X_{\rm max}$ 
on $k$-factor determination}

\begin{figure}[!htb]
\centerline{
\includegraphics[width=8.5cm]{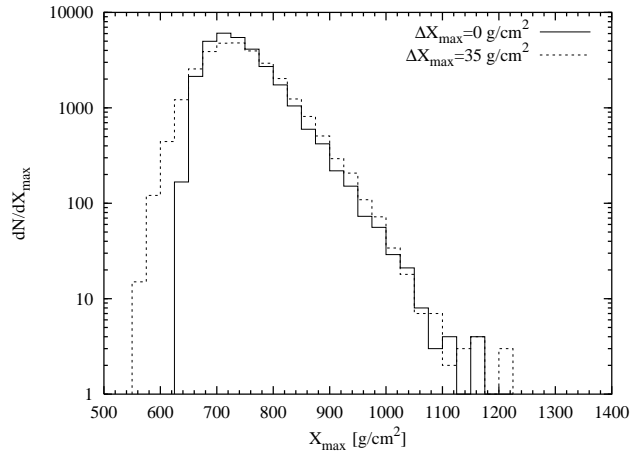}
}
\caption{Distribution of the depth of maximum of proton-induced showers
at energy $E=10^{18}$ eV. 30,000 showers were simulated with 
SIBYLL2.1. The solid histogram is the non-smeared, perfect 
$X_{\rm max}$ resolution distribution, and the dashed histogram
is the solid distribution after smearing it with a Gaussian of standard
deviation $\Delta X_{\rm max}=35~{\rm g/cm^2}$.
\label{fig:xmax-resolution}}
\end{figure}

So far we have assumed an ideal experiment which is able to 
measure the depth of maximum with infinite resolution i.e., 
$\Delta X_{\rm max}=0$. However in the real world the accuracy
is not infinite, in fact the HiRes collaboration has published 
a value of $\Delta X_{\rm max}\sim 35~{\rm g/cm^2}$ for the resolution
in the depth of maximum \cite{HiRes03a,HiRes03b}. The purpose of 
this subsection is to 
explore the effect of the finite resolution on the numerical
value of the $k$ factor. For this purpose we take the $X_{\rm max}$
distribution obtained before assuming a perfect resolution, and we smear
it with a Gaussian of standard deviation equal to the $X_{\rm max}$ resolution
reported by HiRes. An example of the effect of this smearing on the 
original distribution is shown in 
Fig.~\ref{fig:xmax-resolution}. 
As expected the smeared
distribution is wider than the original one 
(by about 15 ${\rm g/cm^2}$). However 
the slope of the distribution is not significantly modified, and as 
a consequence the change in the $k$ factor is still within its 
estimated statistical error.
This is demonstrated in table \ref{tab:k_resol}  
where the $k$ factors obtained from the non-smeared
and the smeared distributions are presented for SIBYLL and QGSjet.

\begin{table*}
\caption{Standard deviation of the $\Delta X=X_{\rm max}-X_{\rm int}$
distribution, $k$ factors and reduced $\chi^2$ of the fits performed
in order to obtain $k$. These quantities are shown for a non-smeared
perfect $X_{\rm max}$ resolution distribution, and for the same distribution
after smearing it with a Gaussian error $\Delta X_{\rm max}=35~{\rm g/cm^2}$.
The results of SIBYLL 2.1 and QGSjet01 hadronic interaction models are shown.
Errors, where shown, are statistical. Only the trailing edge of the 
distribution $100~{g/cm^2}$ beyond shower maximum is used to make the fits.
\label{tab:k_resol}}
\renewcommand{\arraystretch}{1.5}
\begin{tabular}{ c | c c | c c | c c | c c } \hline\hline

Model & \multicolumn{4}{c|}{SIBYLL 2.1} & \multicolumn{4}{c }{QGSjet01} \\ \hline 

$\Delta X_{\rm max}$~& \multicolumn{2}{c|}{$0~{\rm g/cm^2}$} & \multicolumn{2}{c|}{$35~{\rm g/cm^2}$} & \multicolumn{2}{c|}{$0~{\rm g/cm^2}$} & \multicolumn{2}{c }{$35~{\rm g/cm^2}$}  \\ \hline

~E [eV]~&~~$k$~~&~~$\chi^2/{\rm dof}$~~&~~$k$~~&~~$\chi^2/{\rm dof}$~~&~~$k$~~&~~$\chi^2/{\rm dof}$~~&~~$k$~~&~~$\chi^2/{\rm dof}$~~\\ \hline

$10^{18}$~&~$1.15 \pm 0.03$~&~1.24~&~$1.18 \pm 0.03$~&~1.80~&~$1.18 \pm 0.03$~&~1.83~& ~$1.19 \pm 0.02$~&~0.76~\\

$10^{19}$~&~$1.16 \pm 0.02$~&~1.08~&~$ 1.18 \pm 0.02$~&~1.06~&~$1.18 \pm 0.02$~&~0.90~&~$1.17 \pm 0.02$~&~0.75~\\

$10^{20}$~&~$1.14 \pm 0.03$~&~0.94~&~$ 1.20 \pm 0.03$~&~1.62~&~$1.14 \pm 0.02$~&~0.79~&~$1.15 \pm 0.02$~&~0.50~\\ \hline\hline

\end{tabular}
\end{table*}


\begin{figure}[!htb]
\centerline{
\includegraphics[width=10cm]{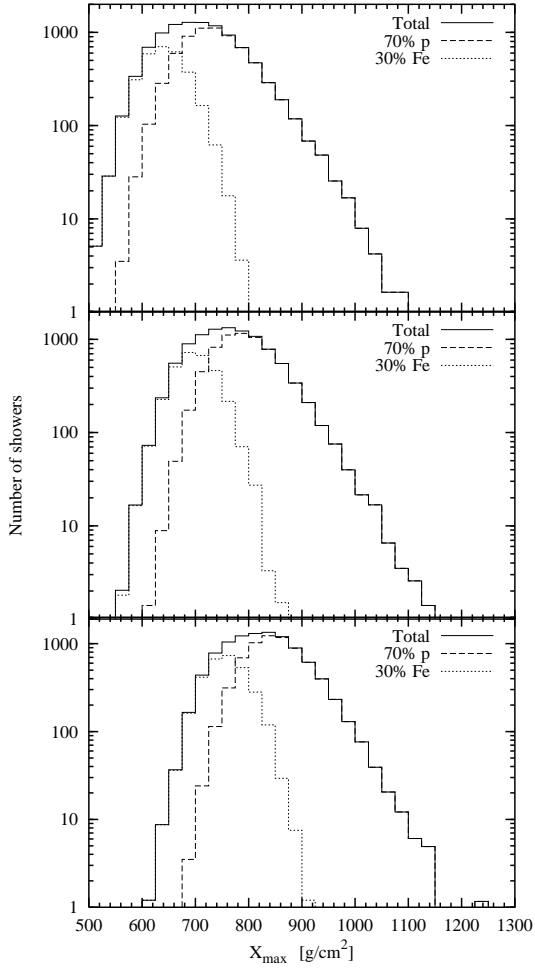}
}
\vspace*{-0.5cm}
\caption{
Depth of maximum distributions for a composition consistent
on $70\%$ protons and $30\%$ iron nuclei. From top to bottom
panels the primary energy is $E=10^{18},~10^{19}$ and $10^{20}$ eV.
The distributions were obtained using the SIBYLL 2.1 hadronic generator. 
A $X_{\rm max}$ Gaussian resolution of $35~{\rm g/cm^2}$ was folded in.
\label{fig:xmax_resol_compos_SIB21}
}
\end{figure}

\subsection{Influence of composition on $k$-factor determination}

A second complication in the determination of the inelastic p-air 
cross section arises due to the mass composition of the primary
cosmic rays. Contamination of the proton spectrum by heavier elements 
may lead to changes in the measured parameter $\Lambda$ and hence a
misinterpretation of the data in terms of the cross section or $k$ factor.
To investigate this issue we simulated iron showers and contaminated the 
proton spectrum assuming the primary cosmic ray composition reported 
by the HiRes collaboration consisting on $70\%$ protons and $30\%$ 
iron \cite{HiRes03a}. 
We assumed that the composition is energy-independent in the energy
range between $10^{18}$ eV and $10^{20}$ eV, as indicated by the HiRes data 
\cite{HiRes03a}.

\begin{figure}[!htb]
\centerline{
\includegraphics[width=10cm]{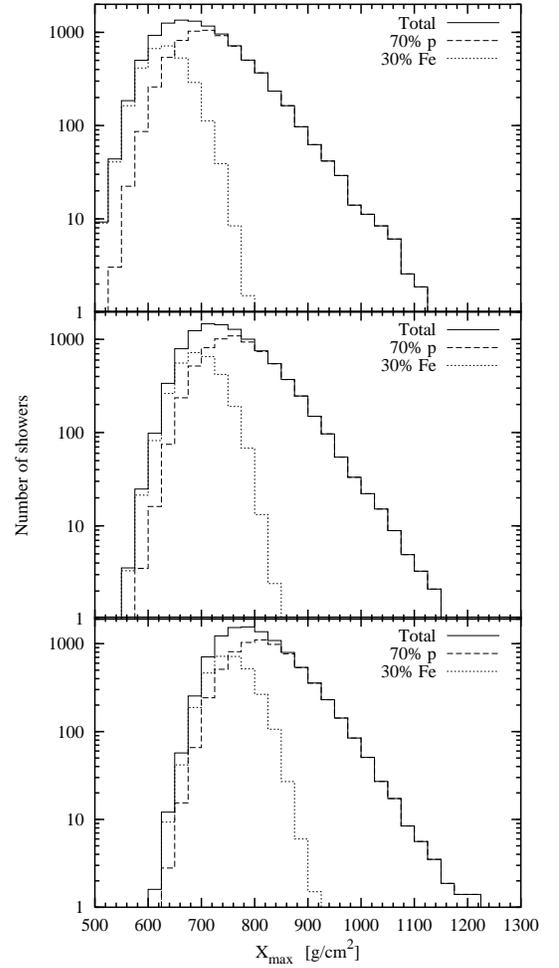}
}
\vspace*{-0.5cm}
\caption{
Same as Fig.~\ref{fig:xmax_resol_compos_SIB21} for QGSjet01.
\label{fig:xmax_resol_compos_QGS01}
}
\end{figure}

Figs.~\ref{fig:xmax_resol_compos_SIB21},~\ref{fig:xmax_resol_compos_QGS01} show
the distributions of $X_{\rm max}$ for the measured HiRes composition,
as predicted by the models SIBYLL 2.1 and QGSjet01 respectively. 
A Gaussian $X_{\rm max}$ resolution of standard deviation 
$\Delta X_{\rm max}=35~{\rm g/cm^2}$ was folded in both the 
proton and the iron distributions. The figures also show the individual
contributions. As is clearly seen, a fraction of $30\%$ of iron nuclei 
does not significantly contribute to the total distribution beyond the peak. 
In fact the change of slope with respect
to a pure protonic composition is very small, almost negligible in 
the tail of the distribution $100~{\rm g/cm^2}$ beyond the peaks 
of the distributions. This conclusion applies for both SIBYLL and 
QGSjet.  As a consequence the $k$ factors do not change with 
respect to those given in table~\ref{tab:k_resol} for protons with 
a resolution in $X_{\rm max}$ of $35~{\rm g/cm^2}$.

\begin{figure}[!htb]
\centerline{
\includegraphics[width=10cm]{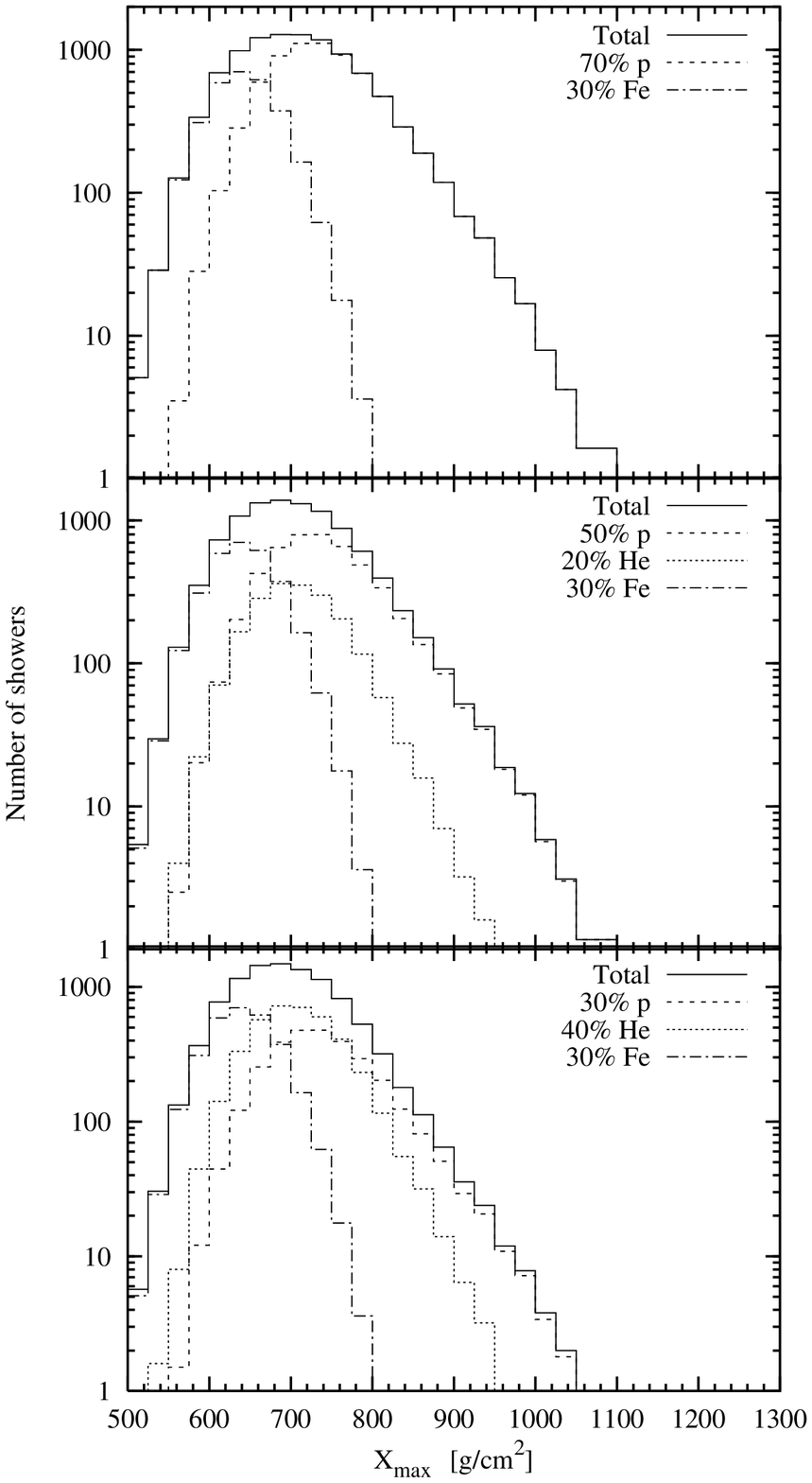}
}
\vspace*{-0.5cm}
\caption{Depth of maximum distributions for a composition consisting
of $70\%$ protons and $30\%$ iron nuclei (top panel), 
$50\%$ protons, $20\%$ helium and $30\%$ iron nuclei (middle panel),
and $30\%$ protons, $40\%$ helium and $30\%$ iron nuclei (bottom panel). 
The primary energy is $E=10^{18}$.
The distributions were obtained using the SIBYLL 2.1 hadronic generator.
A $X_{\rm max}$ Gaussian resolution of $35~{\rm g/cm^2}$ was folded in.
\label{fig:xmax_resol_compos2_SIB21}
}
\end{figure}

Clearly if helium, being the nuclear species that
produces the largest average $X_{\rm max}$, is present in the 
cosmic ray spectrum in this energy range, the contamination might 
be more important and a cut well beyond the peak of the distribution,
in the far tail of the total distribution, has to be performed
in order to avoid a bias in the cross section determination. 
The exact position of the 
cut depends on the fraction of helium in the primary cosmic ray spectrum. 
To further investigate this point we also simulated helium-induced showers
at $E=10^{18}$ eV and we plot the distribution in $X_{\rm max}$ for a 
composition with varying fractions of protons and helium and 
a constant fraction of $30\%$ iron. 
This is shown in Fig.~\ref{fig:xmax_resol_compos2_SIB21}. 
The $k$-factors obtained
when fitting the three $X_{\rm max}$ distributions in 
fig.~\ref{fig:xmax_resol_compos2_SIB21} using only the region
$100~{\rm g/cm^2}$ beyond its peak, are: 
$k=1.19 \pm 0.02,~1.15 \pm 0.02$ and $1.07 \pm 0.02$ for fractions
of proton+helium $70\%+0\%,~50\%+20\%~$ and $30\%+40\%$ 
respectively, keeping the fraction of iron constant. Clearly, if no cut 
is applied at $X_{\rm max}$ larger than the nominal value of 
$100~{\rm g/cm^2}$ beyond the peak of the distribution,  
an important systematic bias is introduced. This will 
induce an error in the determination of the cross section 
if there is a relatively large ($>20\%$) fraction of helium nuclei
in the primary cosmic ray spectrum.


\section{Summary and conclusions\label{sec:dis}}

We have used a hybrid simulation scheme~\cite{Alvarez02} for a 
quantitative evaluation of certain systematic uncertainties in the
interpretation of fluorescence measurements of giant air showers.
The sources of uncertainty we investigate include the model of
hadronic interactions used for shower simulation, the unknown
mixture of primary nuclei in the cosmic radiation
and intrinsic fluctuations in shower
development.  We do not investigate uncertainties in shower reconstruction,
detector acceptance or other technical aspects related to properties of
the environment or performance of the detectors.

As an illustration of uncertainties arising from the need to
extrapolate hadronic interaction models outside the kinematical
region and energies explored by accelerator experiments, we compared two
specific interaction models,  QGSjet~\cite{qgsjet}
and SIBYLL~\cite{Engel99}, both of which agree with each
other and with a range of accelerator data for $\sqrt{s}\sim$~TeV
and below.  We find that the correction for unseen energy (i.e.
energy lost to neutrinos and muons that reach the ground) is
consistently larger for QGSjet than for SIBYLL.  The difference
is such that, for a given track length integral, the assigned
energy will be about 5\% higher when the same data are interpreted
with QGSjet rather than with SIBYLL.

For primary energies below about $10^{19}$~eV/nucleus the fraction
of unseen energy depends significantly on the mass of the primary nucleus.
For example, at $10^{17}$~eV with SIBYLL about 7\% of the primary
energy is lost to neutrinos and muons as compared to 13\% for iron.

In view of the steep cosmic-ray energy spectrum, knowledge of the
energy resolution is of great importance.  The
track length integral can be used to assign energy
when a sufficient portion of
the profile is measured to fix the parameters needed to complete
the integral.  It has an intrinsic resolution of 2-4\%, depending
somewhat on interaction model and energy (narrower at higher energy).
Size of shower at maximum gives only a marginally broader
energy resolution (3-5\%) and can be used when much of the
profile after maximum is not measured, provided care is taken
to correct for a slightly asymmetric distribution.   There is in
both methods some dependence on primary mass of the relation
between the measured quantity and the primary energy which
leads to a $\sim 5$\% systematic uncertainty if the primary
mass is not separately determined.

Intrinsic fluctuations in shower development (after the first 
interaction) affect the relation between the interaction 
length ($\lambda_{\rm int}$) and the slope $\Lambda$
that describes the exponential tail of the $X_{\rm max}$ distribution.
The relation is often expressed with a `$k$ factor´ as
$\Lambda\,=\,k\times\lambda_{\rm int}$.  Differences in $k$ factors
for the range of models studied here are at the level of 5-7\%,
implying a similar uncertainty in the p-air cross section 
that may be inferred from measurements of shower profiles.
Further uncertainties arise to the extent that an unknown
fraction of helium and other nuclei contaminate the tail 
of the measured $X_{\rm max}$ distribution.

\noindent
{\bf Acknowledgments} 
J.A. Ortiz is supported by CNPq/Brasil
and acknowledges the Bartol Research
Institute for its hospitality. This research is supported
in part by NASA Grant NAG5-10919.
RE, TKG \& TS are also supported by the US Department of Energy contract
DE-FG02 91ER 40626. JA-M is also supported by MCyT (FPA 2001-3837 and
FPA 2002-01161). 
The simulations presented here were performed on
Beowulf clusters funded by NSF grant ATM-9977692.

\end{document}